\documentstyle[preprint,revtex]{aps}
\def\ltap{\raisebox{-.4ex}{\rlap{$\sim$}} \raisebox{.4ex}{$<$}}
\begin{document}
\thispagestyle{empty}
\font\fortssbx=cmssbx10 scaled \magstep2
\hbox{ %$\vcenter{\special{insert $disk1:[pheno.tex.inputs]uwlogo.imp}}$
%\hskip.2in $\vcenter{
\fortssbx University of Wisconsin Madison} %$ }
\vspace{.3in}
\hfill\vbox{\hbox{\bf MAD/PH/722}
	    \hbox{October 1992}}\par
\vspace{.2in}
\begin{title}Universal Evolution of CKM Matrix Elements
\end{title}
\author{V.~Barger,  M.~S.~Berger, and
P.~Ohmann}
\begin{instit}
Physics Department, University of Wisconsin, Madison, WI 53706, USA\\
\end{instit}
\begin{abstract}
\baselineskip=36pt %% doublespace abstract
\nonum\section{abstract}
We derive the two-loop evolution equations for the Cabibbo-Kobayashi-Maskawa
matrix. We show that to leading order in the mass and CKM hierarchies the
scaling of the mixings $|V_{ub}|^2$, $|V_{cb}|^2$, $|V_{td}|^2$, $|V_{ts}|^2$
and of the rephase-invariant CP-violating parameter $J$ is universal to all
orders in perturbation theory. In leading order the other CKM elements do not
scale. Imposing the constraint $\lambda _b=\lambda _{\tau}$ at the GUT scale
determines the CKM scaling factor to be $\simeq 0.58$ in the MSSM.
\end{abstract}

\newpage
The weak interaction quark eigenstates and the quark mass eigenstates differ
in the Standard Model as described by the Cabibbo-Kobayashi-Maskawa (CKM)
matrix. In this paper we
show that the scaling of the CKM matrix follows a universal pattern to leading
order in the mass and CKM hierarchies; namely, the CKM mixing elements that
involve the third generation and CP violation scale together, while the other
components of the CKM matrix do not scale to leading order. This makes it
much simpler to consider the form of the quark mixings at any other scale, in
particular at the scale of a grand unified theory (GUT). The common scaling is
a model-independent feature of the evolution, but the amount of scaling can
vary between theories.

The Yukawa matrices ${\bf U}$ and ${\bf D}$ can be diagonalized by biunitary
transformations
\begin{eqnarray}
{\bf U^{diag}}&=&V_u^L{\bf U}V_u^{R\dagger} \;, \\
{\bf D^{diag}}&=&V_d^L{\bf D}V_d^{R\dagger} \;.
\end{eqnarray}
The CKM matrix is then given by
\begin{equation}
V\equiv V_u^LV_d^{L\dagger } \;.
\end{equation}
The Yukawa matrices evolve with energy scale as determined by renormalization
group equations (RGE). This in turn
determines an evolution equation for the ``running'' CKM matrix $V(\mu )$.

The renormalization group scaling to leading order in the mass and CKM
hierarchies can be represented schematically in the following way:
\begin{equation}
{\bf U^{diag}}(M_G^{}) = \left( \begin{array}{c@{\quad}c@{\quad}c}
S_u(\mu )\lambda _u(\mu ) & 0 & 0 \\ 0 &
S_u(\mu )\lambda _c(\mu ) & 0 \\
0 & 0 & S_t(\mu )\lambda _t(\mu )
\end{array} \right) \;,
\end{equation}
\begin{equation}
{\bf D^{diag}}(M_G^{}) = \left( \begin{array}{c@{\quad}c@{\quad}c}
S_d(\mu )\lambda _d(\mu ) & 0 & 0 \\ 0 &
S_d(\mu )\lambda _s(\mu ) & 0 \\
0 & 0 & S_b(\mu )\lambda _b(\mu )
\end{array} \right) \;,
\end{equation}
\begin{equation}
{\bf E^{diag}}(M_G^{}) = \left( \begin{array}{c@{\quad}c@{\quad}c}
S_e(\mu )\lambda _e(\mu ) & 0 & 0 \\ 0 &
S_e(\mu )\lambda _{\mu}(\mu ) & 0 \\
0 & 0 & S_{\tau}(\mu )\lambda _{\tau}(\mu )
\end{array} \right) \;,
\end{equation}
\begin{equation}
{\bf |V|}^2(M_G^{}) = \left( \begin{array}{c@{\quad}c@{\quad}c}
|V_{ud}|^2(\mu )& |V_{us}|^2(\mu ) & S(\mu )|V_{ub}|^2(\mu ) \\
|V_{cd}|^2(\mu ) & |V_{cs}|^2(\mu ) & S(\mu )|V_{cb}|^2(\mu ) \\
S(\mu )|V_{td}|^2(\mu ) & S(\mu )|V_{ts}|^2(\mu ) & |V_{tb}|^2(\mu )
\end{array} \right) \;,
\label{schematic}
\end{equation}
where the scale $\mu $ is the range $m_t\leq \mu \leq M_G^{}$ with $M_G^{}$
the GUT scale. The CP-violating rephase invariant parameter $J$\cite{Jarlskog}
also scales as $J(M_G^{})=S(\mu )J(\mu )$ to leading order.
We have defined
our scaling factors to be unity at the GUT scale, but one could equally well
choose any convenient scale.

The two light generation quark and lepton Yukawa couplings evolve in a common
manner determined by the gauge couplings and traces of the Yukawa matrices,
while the third generation Yukawa couplings receive additional
Yukawa contributions.
This implies that the ratios $\lambda _u/\lambda _c$, $\lambda _d/\lambda _s$,
$\lambda _e/\lambda _{\mu}$ do not evolve.
The scaling pattern in Eq.~(\ref{schematic}) violates unitarity of $V$, but
only at subleading order. For example the relation
$|V_{ud}|^2+|V_{us}|^2+|V_{ub}|^2=1$ is violated by terms that are neglected
to leading order in the evolution of $|V_{ud}|^2$ and $|V_{us}|^2$.
The elements $|V_{ud}|^2$ and $|V_{us}|^2$ must evolve to subleading order to
preserve unitarity. A practical
strategy is to evolve the small mixings $X=|V_{ub}|^2$, $Y=|V_{us}|^2$,
$Z=|V_{cb}|^2$, and $J$ which completely determine the other entries in
the CKM matrix.

In terms of $t=\ln (\mu /M_G^{})$ the two-loop RGEs can be written as
\begin{eqnarray}
{{d{\bf U}}\over {dt}}={1\over {16\pi ^2}}
\Bigg[&\Big (&x_1{\bf I}+x_2{\bf UU^{\dagger }}+a_u{\bf DD^{\dagger }}
\Big )
+{1\over {16\pi ^2}}\Bigg (x_3{\bf I}
+x_4{\bf UU^{\dagger }} \nonumber \\
&+&x_5({\bf UU^{\dagger }})^2+b_u{\bf DD^{\dagger }}+c_u({\bf DD^{\dagger }})^2
+d_u{\bf UU^{\dagger }}{\bf DD^{\dagger }}
+e_u{\bf DD^{\dagger }}{\bf UU^{\dagger }}
\Bigg )\Bigg ]{\bf U} \nonumber \;, \\
\label{dUdt}
\end{eqnarray}
\begin{eqnarray}
{{d{\bf D}}\over {dt}}={1\over {16\pi ^2}}
\Bigg[&\Big (&x_6{\bf I}+x_7{\bf DD^{\dagger }}+a_d{\bf UU^{\dagger }}
\Big )
+{1\over {16\pi ^2}}\Bigg (x_8{\bf I}
+x_9{\bf DD^{\dagger }} \nonumber \\
&+&x_{10}({\bf DD^{\dagger }})^2+b_d{\bf UU^{\dagger }}
+c_d({\bf UU^{\dagger }})^2
+d_d{\bf DD^{\dagger }}{\bf UU^{\dagger }}
+e_d{\bf UU^{\dagger }}{\bf DD^{\dagger }}
\Bigg )\Bigg ]{\bf D} \nonumber \;, \\
\label{dDdt}
\end{eqnarray}
\begin{eqnarray}
{{d{\bf E}}\over {dt}}={1\over {16\pi ^2}}
\Bigg[x_{11}{\bf I}+x_{12}{\bf EE^{\dagger }}
+{1\over {16\pi ^2}}\Bigg (x_{13}{\bf I}
+x_{14}{\bf EE^{\dagger }}+x_{15}({\bf EE^{\dagger }})^2
\Bigg )\Bigg ]{\bf E} \;,
\label{dEdt}
\end{eqnarray}
where the coefficients $x_i$, $a_i$, etc. depend upon the particle content of
the theory and are functions of the gauge and Yukawa couplings, i.e.
$a_i=a_i(g_1^2,g_2^2,g_3^2,
{\bf Tr}[{\bf UU^{\dagger }}],{\bf Tr}[{\bf DD^{\dagger }}],
{\bf Tr}[{\bf EE^{\dagger }}])$ and Higgs quartic couplings.
The coefficients $x_i$ do not enter into the running of the CKM
matrix but do influence the diagonal quark Yukawa evolution; only terms
involving a factor of ${\bf DD^{\dagger}}$ can rotate the ${\bf U}$ matrix,
and only terms with a factor of ${\bf UU^{\dagger}}$ can rotate the ${\bf D}$
matrix.
In the minimal supersymmetric standard model (MSSM) the other
coefficients are
\begin{eqnarray}
a_u&=&a_d=1\;, \\
b_u&=&{2\over 5}g_1^2-{\bf Tr}[3{\bf DD^{\dagger }}
+{\bf EE^{\dagger }}]\;, \\
b_d&=&{4\over 5}g_1^2-{\bf Tr}[3{\bf UU^{\dagger }}]\;, \\
c_u&=&c_d=-2\;, \\
d_u&=&d_d=-2\;, \\
e_u&=&e_d=0\;.
\end{eqnarray}
In the Standard Model they are given by
\begin{eqnarray}
a_u&=&a_d=-{3\over 2}\;, \\
b_u&=&=-{43\over 80}g_1^2+{9\over 16}g_2^2-16g_3^2-2\lambda
+{5\over 4}Y_2(S)\;, \\
b_d&=&=-{79\over 80}g_1^2+{9\over 16}g_2^2-16g_3^2-2\lambda
+{5\over 4}Y_2(S)\;, \\
c_u&=&c_d={11\over 4}\;, \\
d_u&=&d_d=-{1\over 4}\;, \\
e_u&=&e_d=-1\;,
\end{eqnarray}
where
\begin{equation}
Y_2(S)={\bf Tr}[3{\bf UU^{\dagger}}+3{\bf DD^{\dagger}}+{\bf EE^{\dagger}}]\;.
\end{equation}
The coefficients $x_i$ can be found in Refs.~\cite{MV,BBO}.

Following Ma, Pakvasa, Sasaki and Babu\cite{MPS,Babu} we find
the CKM evolution equation
\begin{eqnarray}
{{dV_{i\alpha }}\over {dt}}&=&{1 \over {16\pi ^2}}\left [
a_u\sum _{\beta,j\ne i}
{{\lambda _i^2+\lambda _j^2}\over {\lambda _i^2-\lambda _j^2}}
\hat{\lambda} _{\beta}^2
V_{i\beta }V_{j\beta }^*V_{j\alpha }+a_d\sum _{j,\beta \ne \alpha}
{{\lambda _{\alpha }^2+\lambda _{\beta }^2}\over {\lambda _{\alpha }^2
-\lambda _{\beta }^2}}\hat{\lambda} _j^2
V_{j\beta }^*V_{j\alpha }V_{i\beta }\right ] \nonumber \\
&&\qquad \qquad +{1 \over {(16\pi ^2)^2}}\left [\sum _{\beta,j\ne i}
{{2d_u\lambda _i^2\lambda _j^2+e_u(\lambda_i^4+\lambda _j^4)}
\over {\lambda _i^2-\lambda _j^2}}
\lambda _{\beta}^2
V_{i\beta }V_{j\beta }^*V_{j\alpha }\right .\nonumber \\
&&\qquad \qquad \qquad \qquad \left . +\sum _{j,\beta \ne \alpha}
{{2d_d\lambda _{\alpha }^2\lambda _{\beta }^2+e_d(\lambda _{\alpha }^4
+\lambda _{\beta }^4)}
\over {\lambda _{\alpha }^2
-\lambda _{\beta }^2}}\lambda _j^2
V_{j\beta }^*V_{j\alpha }V_{i\beta }\right ]\;, \label{dVckmdt}
\end{eqnarray}
where
\begin{eqnarray}
\hat{\lambda} _{\beta}^2&=&\lambda _{\beta}^2\left (
1+{{b_u+c_u\lambda _{\beta}^2}
\over {16\pi ^2a_u}}\right )\;, \\
\hat{\lambda} _j^2&=&\lambda _j^2\left (
1+{{b_d+c_d\lambda _j^2}
\over {16\pi ^2a_d}}\right )\;.
\end{eqnarray}
Here $i,j,k=u,c,t,\dots;\;
\alpha,\beta,\gamma =d,s,b,\dots$
We henceforth restrict our considerations to the three-generation case.
Defining the four independent quantities
$X=|V_{ub}|^2$, $Y=|V_{us}|^2$, $Z=|V_{cb}|^2$, and the
parameter $J={\rm Im}V_{ud}^{}V_{cs}^{}V_{us}^*V_{cd}^*$
which can completely specify a unitary CKM matrix, the other elements are
given by\cite{Babu}
\begin{eqnarray}
|V_{ud}|^2&=&1-X-Y\;, \label{Vud} \\
|V_{cs}|^2&=&{{[XYZ+(1-X-Y)(1-X-Z)-2K]}\over {(1-X)^2}}\;,\\
|V_{cd}|^2&=&{{[XZ(1-X-Y)+Y(1-X-Z)+2K]}\over {(1-X)^2}}\;,\\
|V_{tb}|^2&=&1-X-Z\;,\\
|V_{ts}|^2&=&{{[XY(1-X-Z)+(1-X-Y)Z+2K]}\over {(1-X)^2}}\;,\\
|V_{td}|^2&=&{{[X(1-X-Y)(1-X-Z)+YZ-2K]}\over {(1-X)^2}}\;, \label{Vtd}
\end{eqnarray}
where
\begin{equation}
K=[XYZ(1-X-Y)(1-X-Z)-J^2(1-X)^2]^{1/2}\;.
\end{equation}
The full evolution equations for $X$, $Y$, $Z$ and $J$ are given in the
appendix.
Keeping only the leading terms in the mass ($\lambda _c/\lambda _t,
\lambda _u/\lambda _c,
\lambda _s/\lambda _b, \lambda _d/\lambda _s  << 1$)
and CKM ($X, Z, J << 1$) hierarchies, these equations
simplify considerably\cite{BBO} and a universal scaling is found
\begin{equation}
{{dW_1}\over {dt}}=-{{W_1}\over {8\pi ^2}}
\Bigg [\left (a_d\hat{\lambda} _t^2
+a_u\hat{\lambda} _b^2\right )+{{1}\over {(16\pi ^2)}}(e_d+e_u)
\lambda _t^2\lambda _b^2\Bigg ] \;, \label{dW1dt}
\end{equation}
where $W_1=|V_{cb}|^2, |V_{ub}|^2, |V_{ts}|^2, |V_{td}|^2, J$ and
\begin{equation}
{{dW_2}\over {dt}}=0\;, \label{dW2dt}
\end{equation}
where $W_2=|V_{us}|^2, |V_{cd}|^2, |V_{tb}|^2, |V_{cs}|^2, |V_{ud}|^2$.
One does not need the mixing between the first two generations to be small
($Y<<1$) which makes the universality an especially good approximation. To
leading order it is only necessary to include the third generation Yukawa
couplings in $\hat{\lambda }_t^2$ and $\hat{\lambda }_b^2$.
Notice that Eqs.~(\ref{dW1dt})-(\ref{dW2dt})
violate unitarity of $V$, but only at subleading order.
The solution of Eq.~(\ref{dW1dt}) is
\begin{equation}
W_1(M_G^{})=W_1(\mu )S(\mu )\;,
\end{equation}
where $S$ is a scaling factor defined by
\begin{equation}
S(\mu ) = \exp \left \{ -{1\over {8\pi ^2}}\int\limits_{\mu}^{M_G}
\Bigg [\left (a_d\hat{\lambda} _t^2
+a_u\hat{\lambda} _b^2\right )+{{1}\over {(16\pi ^2)}}(e_d+e_u)\lambda _t^2
\lambda _b^2\Bigg ]
d\ln\mu' \right \}
\label{defS} \;.
\end{equation}
This reduces\cite{BBO,DHR} to the scaling factor $y^2(\mu )$ in the one-loop
semianalytic treatment (neglecting $\lambda _b$ and $\lambda _{\tau}$), with
\begin{equation}
y(\mu ) = \exp \left \{ -{1\over {16\pi ^2}}\int\limits_{\mu}^{M_G}
a_d\lambda _t^2(\mu')d\ln\mu' \right \} \label{defy} \;.
\end{equation}
The general behavior of $S(\mu )$
is determined by the sign of $a_d$ (and perhaps also
$a_u$ in models where $\tan \beta$ is large).
In the Standard Model the scaling
factors are greater than one since the one-loop coefficients $a_u$ and $a_d$
are negative.

One might naively have expected there to be contributions to the scaling of
the $W_1$ that are not proportional to $W_1$; for example a contribution to
the running of $|V_{ub}|^2$ of
the form $\lambda _c^2|V_{cb}|^2$ on the right hand side of Eq.~(\ref{dW1dt})
can be of the same order as the contribution $\lambda _t^2|V_{ub}|^2$. We
conclude that no such terms arise.

We find the following RGEs for the Yukawa couplings
\begin{eqnarray}
{{d\lambda _i}\over {dt}}&=&{{\lambda _i}\over {16\pi ^2}}
\Bigg [x_1+x_2\lambda _i^2+
a_u\sum_ {\alpha}\lambda _{\alpha}^2|V_{i\alpha }|^2\nonumber \\
&&+{1\over {16\pi ^2}}\Bigg (x_3+x_4\lambda _i^2+x_5\lambda _i^4
+\sum _{\alpha}\Big (b_u\lambda _{\alpha}^2+c_u\lambda _{\alpha}^4
+(d_u+e_u)\lambda _i^2\lambda _{\alpha}^2\Big )
|V_{i\alpha }|^2\Bigg )\Bigg ]\;, \\
{{d\lambda _{\alpha}}\over {dt}}&=&{{\lambda _{\alpha}}\over {16\pi ^2}}
\Bigg [x_6
+x_7\lambda _{\alpha }^2+
a_d\sum_ i\lambda _i^2|V_{i\alpha }|^2\nonumber \\
&&+{1\over {16\pi ^2}}\Bigg (x_8+x_9\lambda _{\alpha }^2+x_{10}
\lambda _{\alpha }^4
+\sum _i\Big (b_d\lambda _i^2+c_d\lambda _i^4
+(d_d+e_d)\lambda _{\alpha }^2\lambda _i^2\Big )
|V_{i\alpha }|^2\Bigg )\Bigg ]\;, \\
{{d\lambda _a}\over {dt}}&=&{{\lambda _a}\over {16\pi ^2}}
\Bigg [x_{11}+x_{12}\lambda _a^2
+{1\over {16\pi ^2}}\Bigg (x_{13}+x_{14}\lambda _a^2+x_{15}\lambda _a^4
\Bigg )\Bigg ]
\;,
\end{eqnarray}
where $a=e, \mu ,\tau $.
Including only the third generation in the sums, these equations
reduce to the leading order
expressions for $\lambda _t$, $\lambda _b$, $\lambda _{\tau}$ yielding
\begin{eqnarray}
S_t(\mu ) &=& \exp \Bigg \{ {1\over {16\pi ^2}}\int\limits_{\mu}^{M_G}
\Bigg [x_1+x_2\lambda _t^2+
a_u\lambda _b^2\nonumber \\
&&+{1\over {16\pi ^2}}\Bigg (x_3+x_4\lambda _t^2+x_5\lambda _t^4
+\Big (b_u\lambda _b^2+c_u\lambda _b^4
+(d_u+e_u)\lambda _t^2\lambda _b^2\Big )
\Bigg )\Bigg ]d\ln\mu' \Bigg \} \;, \\
S_b(\mu ) &=& \exp \Bigg \{ {1\over {16\pi ^2}}\int\limits_{\mu}^{M_G}
\Bigg [x_6+x_7\lambda _b^2+
a_d\lambda _t^2\nonumber \\
&&+{1\over {16\pi ^2}}\Bigg (x_8+x_9\lambda _b^2+x_{10}\lambda _b^4
+\Big (b_d\lambda _t^2+c_d\lambda _t^4
+(d_d+e_d)\lambda _b^2\lambda _t^2\Big )
\Bigg )\Bigg ]d\ln\mu' \Bigg \} \;, \\
S_{\tau }(\mu ) &=& \exp \Bigg \{ {1\over {16\pi ^2}}\int\limits_{\mu}^{M_G}
\Bigg [x_{11}+x_{12}\lambda _{\tau }^2
+{1\over {16\pi ^2}}\Bigg (x_{13}+x_{14}\lambda _{\tau }^2
+x_{15}\lambda _{\tau }^4\Bigg )\Bigg ]d\ln\mu' \Bigg \} \;,
\end{eqnarray}
respectively. For the first and second generations the corresponding
expressions are
\begin{eqnarray}
S_u(\mu ) &=& \exp \left \{ {1\over {16\pi ^2}}\int\limits_{\mu}^{M_G}
\Bigg [x_1+{1\over {16\pi ^2}}x_3\Bigg ]d\ln\mu' \right \} \;, \\
S_d(\mu ) &=& \exp \left \{ {1\over {16\pi ^2}}\int\limits_{\mu}^{M_G}
\Bigg [x_6+{1\over {16\pi ^2}}x_8\Bigg ]d\ln\mu' \right \} \;, \\
S_e(\mu ) &=& \exp \left \{ {1\over {16\pi ^2}}\int\limits_{\mu}^{M_G}
\Bigg [x_{11}+{1\over {16\pi ^2}}x_{13}\Bigg ]d\ln\mu' \right \} \;.
\end{eqnarray}

In Figure 1 we show contours of constant $S(m_t)$
in the MSSM versus the values
of the Yukawa couplings $\lambda _t$ and $\lambda _b$ at scale $m_t$ and also
at the GUT
scale. The contribution to the scaling from $\lambda _t$ can traded off
against the contribution for $\lambda _b$ as indicated by Eq.~(\ref{defS}).
These contours are shown
versus $m_t$ and $\tan \beta$ in Figure 2. The $m_b(m_b)=4.4$ GeV contour,
which requires $\lambda _b(M_G^{})=\lambda _{\tau}(M_G^{})$,
is plotted as well. The evolution equation for
$R_{b/{\tau}}\equiv \lambda _b/\lambda _{\tau}$ at one-loop
in the MSSM is given by
\begin{equation}
{{dR_{b/\tau}}\over {dt}}={{R_{b/\tau}}\over {16\pi ^2}}
\left (-\sum d_ig_i^2+\lambda _t^2
+3\lambda _b^2-3\lambda _{\tau}^2\right )\;.
\label{dRdt}
\end{equation}
where $d_i=(-4/3,0,16/3)$. For small $\tan \beta $ the bottom and tau Yukawa
couplings can be neglected, and the scaling of $R_{b/\tau }$ factorizes into
scaling due to the gauge couplings and the scaling factor $S$ due to the top
Yukawa. Given a fixed gauge sector scaling, the $m_b$ contours and the
contours of constant $S$ coincide for small $\tan \beta $.
Note that $m_b\simeq 4.4$ GeV implies $S(m_t)\simeq 0.58$.

The numerical calculations performed here are similar to those described in
Ref.~\cite{BBO}. The input values are $\alpha _1(M_Z)^{-1}=58.89$,
$\alpha _2(M_Z)^{-1}=29.75$, and $\alpha _3(M_Z)=0.116$
for the gauge couplings and
$m_b(m_b)=4.4$ GeV, $m_c(m_c)=1.2$ GeV, $m_s(1 {\rm GeV})=0.15$ GeV,
$m_d({\rm 1 GeV})=0.008$ GeV, $m_u(1 {\rm GeV})=0.005$ GeV for the running
fermion masses. We take the lepton masses to be $m_{\tau }=1.7841$ GeV,
$m_{\mu }=0.10566$ GeV and $m_e=5.1100\times 10^{-4}$ GeV.
The GUT scale $M_G^{}$ is determined as the scale at which
unification of $\alpha _1$ and $\alpha _2$ is achieved.
Given an input value for $\tan \beta$ the input masses and the gauge couplings
determine the Yukawa
couplings at the scale $m_t$. We integrate the two-loop RGEs for the gauge
and Yukawa couplings as well as the evolution equations for $X$, $Y$, $Z$, and
$J$ given in the appendix. Our results are
not sensitive to the values of the first and second generation fermion
masses or to the input CKM magnitudes $|V_{cb}(m_t)|=0.043$,
$|V_{ub}(m_t)|=0.0045$, $|V_{us}(m_t)|=0.221$,  $J(m_t)=1.95\times 10^{-5}$.
For experimentally acceptable
values of the quark masses and CKM matrix elements, the exact scaling as
given by the equations in the appendix differ from the universal behavior
described by Eq.~(\ref{dW1dt}) by \ltap 0.1\%.

A good approximation
for evolving the CKM matrix is to use Eq.~(\ref{dW1dt}) to evolve $|V_{ub}|^2$,
$|V_{cb}|^2$ and $J$ and leave $|V_{us}|^2$ constant as in Eq.~(\ref{dW2dt}).
Then calculate the remaining magnitudes $|V_{i\alpha }|$ using
Eqs.~(\ref{Vud})-(\ref{Vtd}).

One can show that the universal scaling behavior described by
Eqs.~(\ref{dW1dt})-(\ref{dW2dt}) is maintained
to all orders in perturbation theory.
However the quantitative effects of the
three-loop contribution are generally smaller
than the sub-leading contributions in the mass and CKM hierarchies.

We now give a sketch of a proof of
the universal behavior at an arbitrary order in
perturbation theory. A higher order contribution will be of the general form
\begin{eqnarray}
{{d{\bf U}}\over {dt}}= {1\over {(16\pi ^2)^q}}
&\Bigg [&f_u^{mn\dots op}({\bf DD^{\dagger }})^m({\bf UU^{\dagger }})^n\dots
({\bf UU^{\dagger }})^o({\bf DD^{\dagger }})^p\Bigg ]{\bf U}+\dots \;,
\label{qloop}
\end{eqnarray}
where $q\geq m+n+\dots +o+p$ represents the loop order. There is an analogous
contribution to $d{\bf D}/dt$.
The exponents $m$ and $p$ could be zero. The coefficient $f_u^{mn\dots op}$ is
calculable in perturbation theory but can be obtained only with tedious
effort; it is a function of the gauge couplings $g_i$ and the sum of the
eigenvalues of the Yukawa couplings matrices, ${\bf Tr}[{\bf UU^{\dagger}}]$,
${\bf Tr}[{\bf DD^{\dagger}}]$, ${\bf Tr}[{\bf EE^{\dagger}}]$, and possibly
other couplings like the quartic Higgs coupling in the standard model.
The term in Eq.~(\ref{qloop}) generates a new contribution
to Eq.~(\ref{dVckmdt}),
\begin{eqnarray}
{{dV_{i\alpha }}\over {dt}}= {1 \over {(16\pi ^2)^q}}&\Bigg [&
f_u^{mn\dots op}\sum _{j\ne i}\Bigg \{
{{1}\over {\lambda _i^2-\lambda _j^2}}\nonumber \\
&&\;\;\;
\sum _{\beta ,k,\gamma,\dots l,\delta }
\Big (\lambda _{\beta}^{2m}\lambda _k^{2n}\dots \lambda _l^{2o}
\lambda _{\delta }^{2p}\lambda _j^2
+\lambda _i^2\lambda _{\beta}^{2p}\lambda _k^{2o}\dots
\lambda _l^{2n}\lambda _{\delta }^{2m}\Big )
V_{i\beta }V_{k\beta }^*V_{k\gamma }V_{l\gamma}^*\dots V_{j\delta}^*
V_{j\alpha }\Bigg \}\Bigg ]+\dots \nonumber \;. \\
\end{eqnarray}
The only terms that contribute to leading order in $dX/dt$, $dY/dt$, $dZ/dt$,
$dJ/dt$ are those in which the indices in the second sum above involve the
third generation
\begin{eqnarray}
{{dV_{i\alpha }}\over {dt}}= {1 \over {(16\pi ^2)^q}}&\Bigg [&
f_u^{mn\dots op}\sum _{j\ne i}\Bigg \{
{{\lambda _i^2+\lambda _j^2}\over {\lambda _i^2-\lambda _j^2}}
\Big (\lambda _b^{2m}\lambda _t^{2n}\dots \lambda _t^{2o}
\lambda _b^{2p}\Big )
V_{ib}V_{tb}^*V_{tb}V_{tb}^*\dots V_{jb}^*
V_{j\alpha }\Bigg \}\Bigg ]+\dots \nonumber \;. \\
\end{eqnarray}
Then to leading order, $|V_{tb}|^2\simeq 1$, and one has
\begin{eqnarray}
{{dV_{i\alpha }}\over {dt}}= {1 \over {(16\pi ^2)^q}}&\Bigg [&
f_u^{mn\dots op}\sum _{j\ne i}\Bigg \{
{{\lambda _i^2+\lambda _j^2}\over {\lambda _i^2-\lambda _j^2}}
\Big (\lambda _t^{2(n+\dots +o)}
\lambda _b^{2(m+\dots +p)}\Big )
V_{ib}V_{jb}^*V_{j\alpha }\Bigg \}\Bigg ]+\dots\nonumber \;,
\end{eqnarray}
which has the same form as Eq.~(\ref{dVckmdt}). Consequently
\begin{equation}
{{dW_1}\over {dt}}= -{{2W_1}\over {(16\pi ^2)^q}}
\Bigg [f_u^{mn\dots op}\left (\lambda _t^{2(n+\dots +o)}
\lambda _b^{2(m+\dots +p)}\right )\Bigg ]+\dots \;. \label{dW1dtf}
\end{equation}
A similar argument applies to the cases $m=0$ and/or $p=0$.

In summary
we have shown that there is a universal scaling pattern in the evolution of the
CKM matrix when only the leading order terms in the mass and CKM hierarchies
are kept. This is a very good approximation given the observed hierarchy
of the quark masses and CKM matrix elements. This scaling
behavior persists to all orders in perturbation theory. Imposing a GUT scale
constraint $\lambda _b(M_G^{})=\lambda _{\tau}(M_G^{})$ constrains the
amount of scaling. For $m_b(m_b)=4.4$ GeV the scaling factor of the CKM matrix
is $S\simeq 0.58$.

After this work was completed we learned that similar conclusions about
the scaling
of the CKM elements have been obtained at the one-loop level
by Babu and Shafi\cite{BS}.

\acknowledgements
This research was supported
in part by the University of Wisconsin Research Committee with funds granted by
the Wisconsin Alumni Research Foundation, in part by the U.S.~Department of
Energy under contract no.~DE-AC02-76ER00881, and in part by the Texas National
Laboratory Research Commission under grant no.~RGFY9273.  Further support was
also provided by U.S. Department of Education under Award No. P200A80214.
PO was supported in part by an NSF Graduate Fellowship.

\nonum\section{appendix}
The evolution equations for $|V_{i\alpha }|^2$ can be derived from
Eq.~(\ref{dVckmdt}) using the substitutions in Eqs.~(\ref{Vud})-(\ref{Vtd}),
as performed by Babu\cite{Babu} at the one-loop level. At the two-loop level
one obtains
\begin{eqnarray}
{{dX}\over {dt}}&=&{2\over {16\pi ^2}}\Bigg [a_u{{\lambda _u^2+\lambda _c^2}
\over {\lambda _u^2-\lambda _c^2}}
\Bigg \{(\hat{\lambda} _b^2-\hat{\lambda} _d^2)XZ
\nonumber \\
&&+{{(\hat{\lambda} _d^2-\hat{\lambda} _s^2)}\over {1-X}}(XYZ-K)\Bigg \}
\nonumber \\
&&+a_u{{\lambda _u^2+\lambda _t^2}
\over {\lambda _u^2-\lambda _t^2}}
\Bigg \{(\hat{\lambda} _b^2-\hat{\lambda} _d^2)X(1-X-Z)
\nonumber \\
&&+{{(\hat{\lambda} _d^2-\hat{\lambda} _s^2)}\over {1-X}}(XY(1-X-Z)+K)\Bigg \}
\nonumber \\
&&+a_d{{\lambda _b^2+\lambda _s^2}
\over {\lambda _b^2-\lambda _s^2}}
\Bigg \{(\hat{\lambda} _u^2-\hat{\lambda} _t^2)XY
\nonumber \\
&&+{{(\hat{\lambda} _t^2-\hat{\lambda} _c^2)}\over {1-X}}(XYZ-K)\Bigg \}
\nonumber \\
&&+a_d{{\lambda _b^2+\lambda _d^2}
\over {\lambda _b^2-\lambda _d^2}}
\Bigg \{(\hat{\lambda} _u^2-\hat{\lambda} _t^2)X(1-X-Y)
\nonumber \\
&&+{{(\hat{\lambda} _t^2-\hat{\lambda} _c^2)}
\over {1-X}}(XZ(1-X-Y)+K)\Bigg \}\Bigg ]
\nonumber \\
&+&{2\over {(16\pi ^2)^2}}\Bigg [{{2d_u\lambda _u^2\lambda _c^2+e_u
(\lambda _u^4+\lambda _c^4)}
\over {\lambda _u^2-\lambda _c^2}}
\Bigg \{(\lambda _b^2-\lambda _d^2)XZ
\nonumber \\
&&+{{(\lambda _d^2-\lambda _s^2)}\over {1-X}}(XYZ-K)\Bigg \}
\nonumber \\
&&+{{2d_u\lambda _u^2\lambda _t^2+e_u(\lambda _u^4+\lambda _t^4)}
\over {\lambda _u^2-\lambda _t^2}}
\Bigg \{(\lambda _b^2-\lambda _d^2)X(1-X-Z)
\nonumber \\
&&+{{(\lambda _d^2-\lambda _s^2)}\over {1-X}}(XY(1-X-Z)+K)\Bigg \}
\nonumber \\
&&+{{2d_d\lambda _b^2\lambda _s^2+e_d(\lambda _b^4+\lambda _s^4)}
\over {\lambda _b^2-\lambda _s^2}}
\Bigg \{(\lambda _u^2-\lambda _t^2)XY
\nonumber \\
&&+{{(\lambda _t^2-\lambda _c^2)}\over {1-X}}(XYZ-K)\Bigg \}
\nonumber \\
&&+{{2d_d\lambda _b^2\lambda _d^2+e_d(\lambda _b^4+\lambda _d^4)}
\over {\lambda _b^2-\lambda _d^2}}
\Bigg \{(\lambda _u^2-\lambda _t^2)X(1-X-Y)
\nonumber \\
&&+{{(\lambda _t^2-\lambda _c^2)}\over {1-X}}(XZ(1-X-Y)+K)\Bigg \}\Bigg ]
\nonumber \;, \\ \label{dXdt}
\\
\nonumber \\
{{dY}\over {dt}}&=&{2\over {16\pi ^2}}\Bigg [a_u{{\lambda _u^2+\lambda _c^2}
\over {\lambda _u^2-\lambda _c^2}}
\Bigg \{{{(\hat{\lambda} _d^2-\hat{\lambda} _b^2)}\over {1-X}}(XYZ-K)
\nonumber \\
&&+{{(\hat{\lambda} _s^2-\hat{\lambda} _d^2)}\over {(1-X)^2}}
Y(XYZ+(1-X-Y)(1-X-Z)-2K)\Bigg \}
\nonumber \\
&&+a_u{{\lambda _u^2+\lambda _t^2}
\over {\lambda _u^2-\lambda _t^2}}
\Bigg \{{{(\hat{\lambda} _b^2-\hat{\lambda} _d^2)}\over {1-X}}(XY(1-X-Z)+K)
\nonumber \\
&&+{{(\hat{\lambda} _s^2-\hat{\lambda} _d^2)}\over {(1-X)^2}}
Y(XY(1-X-Z)+Z(1-X-Y)+2K)\Bigg \}
\nonumber \\
&&+a_d{{\lambda _s^2+\lambda _b^2}
\over {\lambda _s^2-\lambda _b^2}}
\Bigg \{(\hat{\lambda} _u^2-\hat{\lambda} _t^2)XY
\nonumber \\
&&+{{(\hat{\lambda} _t^2-\hat{\lambda} _c^2)}\over {1-X}}(XYZ-K)\Bigg \}
\nonumber \\
&&+a_d{{\lambda _s^2+\lambda _d^2}
\over {\lambda _s^2-\lambda _d^2}}
\Bigg \{(\hat{\lambda} _u^2-\hat{\lambda} _t^2)Y(1-X-Y)
\nonumber \\
&&+{{(\hat{\lambda} _c^2-\hat{\lambda} _t^2)}\over {(1-X)^2}}
(XYZ(1-X-Y)-Y(1-X-Y)(1-X-Z)-K(1-X-2Y))\Bigg \}\Bigg ] \nonumber \\
&+&{2\over {(16\pi ^2)^2}}\Bigg [{{2d_u\lambda _u^2\lambda _c^2
+e_u(\lambda _u^4+\lambda _c^4)}
\over {\lambda _u^2-\lambda _c^2}}
\Bigg \{{{(\lambda _d^2-\lambda _b^2)}\over {1-X}}(XYZ-K)
\nonumber \\
&&+{{(\lambda _s^2-\lambda _d^2)}\over {(1-X)^2}}
Y(XYZ+(1-X-Y)(1-X-Z)-2K)\Bigg \}
\nonumber \\
&&+{{2d_u\lambda _u^2\lambda _t^2+e_u(\lambda _u^4+\lambda _t^4)}
\over {\lambda _u^2-\lambda _t^2}}
\Bigg \{{{(\lambda _b^2-\lambda _d^2)}\over {1-X}}(XY(1-X-Z)+K)
\nonumber \\
&&+{{(\lambda _s^2-\lambda _d^2)}\over {(1-X)^2}}
Y(XY(1-X-Z)+Z(1-X-Y)+2K)\Bigg \}
\nonumber \\
&&+{{2d_d\lambda _s^2\lambda _b^2+e_d(\lambda _s^4+\lambda _b^4)}
\over {\lambda _s^2-\lambda _b^2}}
\Bigg \{(\lambda _u^2-\lambda _t^2)XY
\nonumber \\
&&+{{(\lambda _t^2-\lambda _c^2)}\over {1-X}}(XYZ-K)\Bigg \}
\nonumber \\
&&+{{2d_d\lambda _s^2\lambda _d^2+e_d(\lambda _s^4+\lambda _d^4)}
\over {\lambda _s^2-\lambda _d^2}}
\Bigg \{(\lambda _u^2-\lambda _t^2)Y(1-X-Y)
\nonumber \\
&&+{{(\lambda _c^2-\lambda _t^2)}\over {(1-X)^2}}
(XYZ(1-X-Y)-Y(1-X-Y)(1-X-Z)-K(1-X-2Y))\Bigg \}\Bigg ]
\nonumber \;, \\
\\
\nonumber \\
{{dZ}\over {dt}}&=&{2\over {16\pi ^2}}\Bigg [a_u{{\lambda _c^2+\lambda _u^2}
\over {\lambda _c^2-\lambda _u^2}}
\Bigg \{(\hat{\lambda} _b^2-\hat{\lambda} _d^2)XZ
\nonumber \\
&&+{{(\hat{\lambda} _d^2-\hat{\lambda} _s^2)}\over {1-X}}(XYZ-K)\Bigg \}
\nonumber \\
&&+a_u{{\lambda _c^2+\lambda _t^2}
\over {\lambda _c^2-\lambda _t^2}}
\Bigg \{(\hat{\lambda} _b^2-\hat{\lambda} _d^2)Z(1-X-Z)
\nonumber \\
&&+{{(\hat{\lambda} _s^2-\hat{\lambda} _d^2)}\over {(1-X)^2}}
(XYZ(1-X-Z)-Z(1-X-Y)(1-X-Z)-K(1-X-2Z))\Bigg \}
\nonumber \\
&&+a_d{{\lambda _b^2+\lambda _s^2}
\over {\lambda _b^2-\lambda _s^2}}
\Bigg \{{{(\hat{\lambda} _u^2-\hat{\lambda} _t^2)}\over {1-X}}(K-XYZ)
\nonumber \\
&&+{{(\hat{\lambda} _c^2-\hat{\lambda} _t^2)}\over {(1-X)^2}}
Z(XYZ+(1-X-Y)(1-X-Z)-2K)\Bigg \}
\nonumber \\
&&+a_d{{\lambda _b^2+\lambda _d^2}
\over {\lambda _b^2-\lambda _d^2}}
\Bigg \{{{(\hat{\lambda} _t^2-\hat{\lambda} _u^2)}\over {1-X}}(XZ(1-X-Y)+K)
\nonumber \\
&&+{{(\hat{\lambda} _c^2-\hat{\lambda} _t^2)}\over {(1-X)^2}}
Z(XZ(1-X-Y)+Y(1-X-Z)+2K)\Bigg \}\Bigg ] \nonumber \\
&+&{2\over {(16\pi ^2)^2}}\Bigg [{{2d_u\lambda _c^2\lambda _u^2
+e_u(\lambda _c^4+\lambda _u^4)}
\over {\lambda _c^2-\lambda _u^2}}
\Bigg \{(\lambda _b^2-\lambda _d^2)XZ
\nonumber \\
&&+{{(\lambda _d^2-\lambda _s^2)}\over {1-X}}(XYZ-K)\Bigg \}
\nonumber \\
&&+{{2d_u\lambda _c^2\lambda _t^2+e_u(\lambda _c^4+\lambda _t^4)}
\over {\lambda _c^2-\lambda _t^2}}
\Bigg \{(\lambda _b^2-\lambda _d^2)Z(1-X-Z)
\nonumber \\
&&+{{(\lambda _s^2-\lambda _d^2)}\over {(1-X)^2}}
(XYZ(1-X-Z)-Z(1-X-Y)(1-X-Z)-K(1-X-2Z))\Bigg \}
\nonumber \\
&&+{{2d_d\lambda _b^2\lambda _s^2+e_d(\lambda _b^4+\lambda _s^4)}
\over {\lambda _b^2-\lambda _s^2}}
\Bigg \{{{(\lambda _u^2-\lambda _t^2)}\over {1-X}}(K-XYZ)
\nonumber \\
&&+{{(\lambda _c^2-\lambda _t^2)}\over {(1-X)^2}}
Z(XYZ+(1-X-Y)(1-X-Z)-2K)\Bigg \}
\nonumber \\
&&+{{2d_d\lambda _b^2\lambda _d^2+e_d(\lambda _b^4+\lambda _d^4)}
\over {\lambda _b^2-\lambda _d^2}}
\Bigg \{{{(\lambda _t^2-\lambda _u^2)}\over {1-X}}(XZ(1-X-Y)+K)
\nonumber \\
&&+{{(\lambda _c^2-\lambda _t^2)}\over {(1-X)^2}}
Z(XZ(1-X-Y)+Y(1-X-Z)+2K)\Bigg \}\Bigg ]
\nonumber \;, \\
\\
\nonumber \\
{{dJ}\over {dt}}&=&{-(J/2) \over {16\pi ^2}}\left [a_u\sum _{\beta,j\ne i}
{{\lambda _i^2+\lambda _j^2}\over {\lambda _i^2-\lambda _j^2}}
\hat{\lambda} _{\beta}^2
(|V_{i\beta }|^2-|V_{j\beta }|^2)+a_d\sum _{j,\beta \ne \alpha}
{{\lambda _{\alpha }^2+\lambda _{\beta }^2}\over {\lambda _{\alpha }^2
-\lambda _{\beta }^2}}\hat{\lambda} _j^2
(|V_{j\alpha }|^2-|V_{j\beta }|^2)
\right ] \nonumber \\
&&+{-(J/2) \over {(16\pi ^2)^2}}\left [\sum _{\beta,j\ne i}
{{2d_u\lambda _i^2\lambda _j^2+e_u(\lambda _i^4+\lambda _j^4)}
\over {\lambda _i^2-\lambda _j^2}}
\lambda _{\beta}^2
(|V_{i\beta }|^2-|V_{j\beta }|^2)\right . \nonumber \\
&&\vspace*{1in}\left . +\sum _{j,\beta \ne \alpha}
{{2d_d\lambda _{\alpha }^2\lambda _{\beta }^2+e_d
(\lambda _{\alpha}^4+\lambda _{\beta}^4)}\over {\lambda _{\alpha }^2
-\lambda _{\beta }^2}}\lambda _j^2
(|V_{j\alpha }|^2-|V_{j\beta }|^2)
\right ] \;. \label{dJdt}
\nonumber \\
\end{eqnarray}
Replacing $\hat{\lambda }_i\rightarrow \lambda _i$,
$\hat{\lambda }_{\alpha }\rightarrow \lambda _{\alpha }$ and omitting the
second contributions proportional to $(16\pi ^2)^{-2}$, one recovers the
one-loop results of Babu\cite{Babu} (our definitions of $X$ and $Z$
differ from Ref.~\cite{Babu}). The
two-loop equations have the same overall structure as the one-loop
equations because both contain the same factor
$V_{i\beta }V_{j\beta }^*V_{j\alpha }$ in Eq.~(\ref{dW1dt}).
Eqs.~(\ref{dXdt})-(\ref{dJdt}),
together with the evolution equations for the gauge couplings $g_i$
and the Yukawa couplings $\lambda _i$, form a coupled set of differential
equations that can be solved numerically. In their full form these equations
together with Eqs.~(\ref{Vud})-(\ref{Vtd})
preserve the unitarity of the CKM matrix to all orders in the hierarchy.

\newpage
\nonum\section{figures}
\vskip 0.5in

\noindent Fig. 1. Contours of constant $S^{1/2}(m_t)$ in the
MSSM are shown for values of
$\lambda _t$ and $\lambda _b$ at (a) $\mu = m_t$ and (b) $\mu = M_G^{}$.
We have taken $\alpha _3(M_Z)=0.116$.
\vskip 0.5in

\noindent Fig. 2.  Contours of constant $S^{1/2}(m_t)$ in the MSSM are
shown in the
$m_t,\tan \beta$ plane for $\alpha _3(M_Z)=0.116$. The dashed line is the
$m_b(m_b)=4.4$ GeV contour obtained from the GUT scale condition
$\lambda _b(M_G^{})=\lambda _{\tau}(M_G^{})$. The {\bf X} marks the spot
at which $\lambda _t(M_G^{})=\lambda _b(M_G^{})=\lambda _{\tau}(M_G^{})$ for
this $m_b$ contour. In the small $\tan \beta $
region a linear relationship exists between $m_t$ and $\sin \beta $ for each
contour.

\end{document}